# Fractal Time Series Analysis of Social Network Activities

Lyudmyla Kirichenko, Vitalii Bulakh
Department of Applied Mathematics
Kharkiv National University of Radioelectronics
Kharkiv, Ukraine
Lyudmyla.kirichenko@nure.ua,
bulakhvitalii@gmail.com

Tamara Radivilova
Department of Infocommunication Engineering
Kharkiv National University of Radioelectronics
Kharkiv, Ukraine
tamara.radivilova@gmail.com

*Abstract*— **In the work, a comparative correlation and fractal analysis of time series of Bitcoin crypto currency rate and community activities in social networks associated with Bitcoin was conducted. A significant correlation between the Bitcoin rate and the community activities was detected. Time series fractal analysis indicated the presence of self-similar and multifractal properties. The results of researches showed that the series having a strong correlation dependence have a similar multifractal structure.**

## I. INTRODUCTION

In recent years, the intellectual capabilities of social networks for financial markets have been discovered. Social networking data is used for commercial purposes to automatically extract customer reviews of products or brands, search for new customers, analyze preferences and influence social networking users in voting, elective, on the market, etc. Financial market analysts for investment decisions rely on traditional sources of information, such as disclosure of information about the companies, market news, reports, and so on. As in the past few years social networks have rapidly penetrated into the various aspects of our lives, and contain a huge amount of information, the social networking data provide an alternative source of information for analysts and financial market investors [1].

Many factors have an impact on the financial markets, and the emergence of cryptocurrency has had a huge impact on them. Bitcoin is a major cryptocurrency and derivative financial instruments, which is very in demand in recent years [2].

Bitcoin is a decentralized payment system of electronic currencies, created in 2008 by Satoshi Nakamoto [3], which has steadily grown to a market capitalization of over 54 trillions US dollars and more than 200 thousands transactions a day. Bitcoin is an IT innovation that uses sophisticated cryptographic techniques to sign transactions and determine control means. Bitcoin is a peer-to-peer system in which users conduct a transaction directly, without the need for an intermediary, and in which transactions are recorded for later verification by all nodes in a public distributed book called Blockchain. In the book Blockchain each transaction is recorded and it provides an excellent source of data that can be analyzed and funds and communications between users can be received.

One of the directions of research data in social networks and the financial sector is to analyze of the relationship between information about Bitcoin and Blockchain contained in social networks, and at the Bitcoin rate.

Now it has become generally accepted that financial markets have fractal properties [4]. In recent years, the analysis of the Bitcoin crypto currency rate for the presence of fractal properties has been carried out [5]. Also, researchers carried out a fractal analysis of the structure of social networks [6] and information flows in social networks [7]. Later, a number of studies were carried out, which showed the potential of social networks in providing financial market participants with valuable information and assistance in forming investment views. Many industrial companies have experimented with new methods of financial analysis by monitoring social dynamics, for example, DataSift (social data company that collects and analyzes unstructured data from social networking sites such as Facebook) and Cayman Atlantic (an investment management company that manages managed trading accounts based on the detection of real-time events with data from social networks using a sentiment analysis).

Predictive analysis of Bitcoin crypto currency rate based on information flows of social networks is considered using content analysis [1], graph theory [8], graph theory and through several different similarity metrics [2], cross-correlation and sentiment analysis [9], linear regression analysis [10,11], textual analysis, econometric models [12]. Social networks and interactions between market participants have a great influence on the Bitcoin rate.

In [13], it is shown that the growing popularity of Bitcoin causes an increase amount of search, which in turn leads to a higher activity in social networks toward to Bitcoin. Buying Bitcoins by users leads to greater interest, which leads to higher prices, which will eventually be returned to amount of search (namely a social feedback cycle).

The purpose of the work is to investigating the relationships and conducting fractal analysis of Bitcoin crypto currency rate time series and community activities in social networks associated with this crypto currency.

## II. CHARACTERISTICS OF SELF-SIMILAR AND MULTIFRACTAL PROCESSES

A stochastic continious-time process $X(t)$ is said to be self-similar of a parameter $H$, $0 < H < 1$, if for any $a > 0$

processes $X(at)$ and $a^{-H}X(at)$ have same finite-dimensional distributions:

$$Law\{X(at)\} = Law\{a^H X(t)\}. \qquad (1)$$

The parameter $H$ is called Hurst exponent. It is a measure of self-similarity or a characteristic of long-range dependence of the process. For values $0.5 < H < 1$ time series behavior is persistent. The value $H = 0.5$ correspond to the absence of any memory about the past. The values $0 < H < 0.5$ imply an antipersistent behavior of time series. The moments of the self-similar random process (1) can be expressed as $E\left[|X(t)^q|\right] \propto t^{qH}$.

Contrasting to the self-similar processes (1), multifractal processes have more complex scaling behavior:

$$Law\{X(at)\} = Law\{M(a) \cdot X(t)\}, \qquad (2)$$

where $M(a)$ is a random function. In case of self-similar process $M(a) = a^H$. [14,15].

For multifractal processes the following relation holds:

$$E\left[|X(t)|^q\right] \propto t^{qh(q)} \qquad (3)$$

where $h(q)$ is generalized Hurst exponent. The value $h(q)$ at $q = 2$ has the same degree of self-similarity $H$. The generalized Hurst exponent of a monofractal process is constant: $h(q)=H$.

As a characteristic of heterogeneity of multifractal time series the range of the generalized Hurst exponent $\Delta h = h(q_{min}) - h(q_{max})$ is used. For monofractal processes the generalized Hurst exponent is a straight line: $h(q)=H$, $\Delta h=0$. This implies that the greater heterogeneity of the process, the greater the range $\Delta h$.

### III. A Method of Multifractal Detrended Fluctuation Analysis

One of the most popular research tools for multifractal analysis of the time series is the method of multifractal detrended fluctuation analysis (MFDFA) [14,16]. It is focused on processing of non-stationary trended series.

According to the MFDFA method, for the initial time series $x(t)$ the cumulative time series $y(t) = \sum_{i=1}^{t} x(i)$ is constructed and then divided into $N$ segments of length $\tau$. Also, for each segment $y(t)$ the following fluctuation function is calculated:

$$F^2(\tau) = \frac{1}{\tau} \sum_{t=1}^{\tau} (y(t) - Y_m(t))^2 \qquad (4)$$

where $Y_m(t)$ is a local $m$-polynomial trend within the given segment. The function $F(\tau)$ averaged on the whole of the time series $y(t)$ has scaling on the segment of length $\tau$:

$$F(\tau) \propto \tau^H$$

In the study multifractal properties the dependence of the following fluctuation function $F_q(\tau)$ on the parameter $\tau$ is considered:

$$F_q(\tau) = \left\{ \frac{1}{N} \sum_{i=1}^{N} [F^2(\tau)]^{\frac{q}{2}} \right\}^{\frac{1}{q}}$$

If the investigated series is multifractal and has a long-term dependence, the fluctuation function $F_q(\tau)$ is represented by a power law

$$F_q(\tau) \propto \tau^{h(q)}$$

where $h(q)$ is generalized Hurst exponent.

### IV. Data

In the work, a comparative fractal analysis was conducted for three community groups in the social network and the Bitcoin crypto currency rate. To conduct research in the Facebook network, three social communities were selected: Bitcoin Product / service of 196113 users, Bitcoin Finance company of 70195 users, group Blockchain of 61805 users. For the period from October 2016 to July 2017, the data of likes, reposts for each group and the Bitcoin crypto currency rate were recorded. The records were analyzed to determine their identifier, date-time representation, type. Thus, seven time series of daily data were formed:

*Bitcoin_exchange* is Bitcoin crypto currency rate relative to USA dollar;
*Bitcoin_likes* is a number of likes in community Bitcoin Product/service;
*Bitcoin_reposts* is a number of reposts in community Bitcoin Product/service;
*BitcoinF_likes* is a number of likes in community Bitcoin Finance company;
*BitcoinF_reposts* is a number of reposts in community Bitcoin Finance company;
*Blockchain_likes* is a number of likes in community Blockchain;
*Blockchain_reposts* is a number of reposts in community Blockchain.

Fig. 1 shows the time series of the values of the *Bitcoin_exchange* in the period from October 17, 2016 to May 30, 2017 and the corresponding series *Bitcoin_like* and *Bockchain_likes*.

Correlation analysis was performed to investigate the relationship of series *Bitcoin_exchange* with social activities of communities associated with it. A significant correlation between the Bitcoin rate and the communities activity was detected. The cros-correlation functions of the *Bitcoin_exchange* and *Bitcoin_likes* series (a strong correlation) and *Bitcoin_exchange* and *blockchain_likes* (a weak correlation with a lag) are presented in Fig. 2.

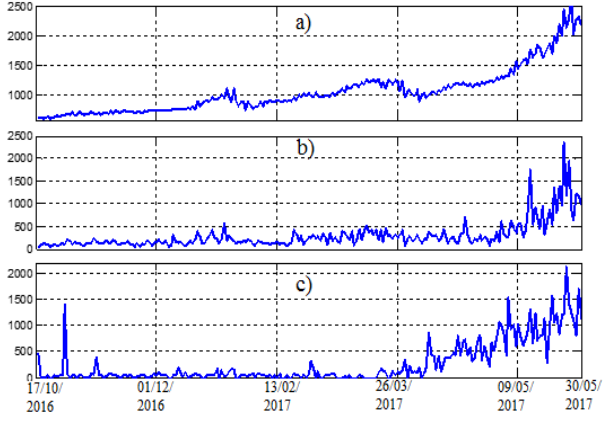

Fig. 1. Time series Bitcoin_exchange, Bitcoin_like and Bockchain_likes

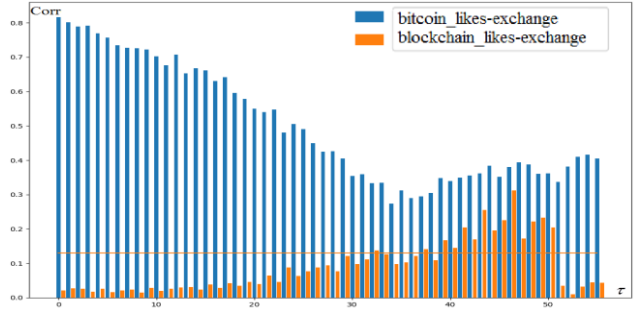

Fig. 2. Cross-correlation function Bitcoin_exchange vs Bitcoin_likes and Bitcoin_exchange vs Blockchain_likes

## V. Results of fractal analysis

Before the multifractal analysis is conducted, it is necessary to investigate the fluctuation function $F(\tau)$. Self-similar behavior corresponds to the presence of linear dependence between $\log F(\tau)$ and $\log(\tau)$. [16]. Fig. 3 shows a graph for the time series *Bitcoin_likes* presented in Fig. 1b. The resulting curve is quite well approximated by a straight line with a slope corresponding to the Hurst exponent.

The research of fluctuation functions for the series of Bitcoin prices and activity groups associated with this cryptocurrency have shown the presence of self-similar properties. Also, the multifractal properties of these series were studied.

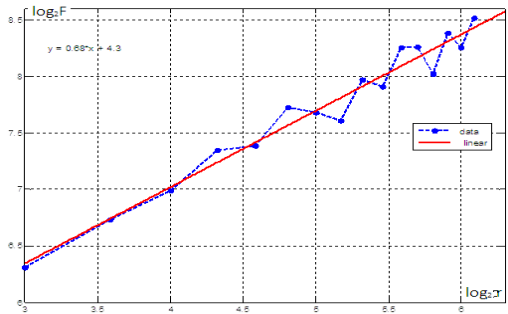

Fig. 3. Dependence of the fluctuation function $\log F(\tau)$ on $\log \tau$

Fig. 4 presents the functions of the generalized Hurst exponent for series *Bitcoin_exchange*, *Bitcoin_likes* and *Blockchain_likes* during the period from October 17, 2016 on May 30, 2017.

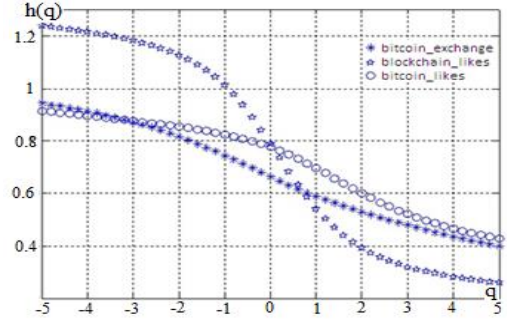

Fig. 4. Generalized Hurst exponents $h(q)$ for *Bitcoin_exchange*, *Bitcoin_likes* and *Blockchain_likes*

The study have shown that the series with a strong correlation dependence have a similar multifractal structure. Correlation analysis of the relationships between the series of Bitcoin exchange and ones of social communities activities (see Fig. 2) showed that the statistical relationship can be of various types. The series *Bitcoin_exchange* and *Bitcoin_likes* have strong correlation and close values of the multifractal characteristic. A weak correlation between *Bitcoin_exchange* and *Blockchain_like* is manifested in the difference of the multifractal structure of these series (see Fig. 4).

The study demonstrated the presence of multifractal properties for all series of community activities. The tabulated values of $h(q)$ at $q = -5, 5$ and values of the Hurst exponent $H$ for the considered series of community activities from October 17, 2016 to July 27, 2017 are given in the table. Since most likely the series of likes and reposts are strongly related to each other, they have close values of the generalized Hurst exponent. The series *Bitcoin_like* and *Bitcoin_repost* from October 17, 2016 to July 27, 2017 are depicted in Fig. 5. Their strong correlation dependence is evident.

TABLE I. Values of multifractal characteristics

|  | $h(q=-5)$ | $h(q=5)$ | $H$ |
|---|---|---|---|
| Bitcoin_likes | 0.78 | 0.42 | 0.54 |
| Bitcoin_reposts | 0.75 | 0.38 | 0.55 |
| BitcoinF_likes | 0.95 | 0.52 | 0.63 |
| BitcoinF_reposts | 1.15 | 0.43 | 0.64 |
| Blockchain_likes | 1.23 | 0.25 | 0.41 |
| Blockchain _reposts | 1.41 | 0.23 | 0.45 |

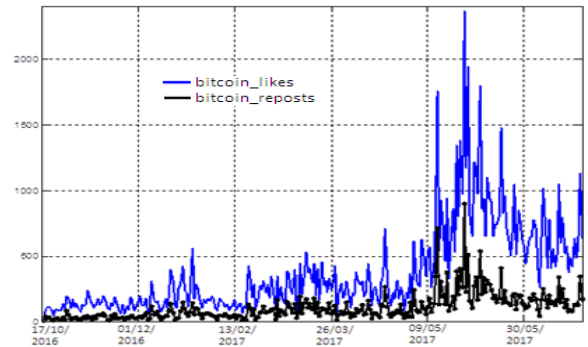

Fig. 5. Time series of likes and repots for one community

Fig. 6 shows the corresponding functions of the generalized Hurst exponent. They have practically the same range of $h(q)$ and close values of the Hurst exponent $H$.

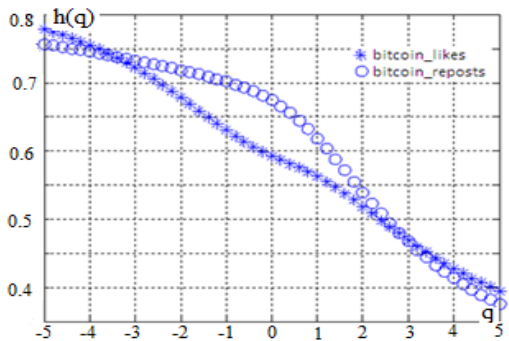

Fig. 6. Generalized Hurst exponents $h(q)$ for *Bitcoin_likes* and *Bitcoin_repost*

Similar results were observed for other investigated series (see Table). The trend is the greater the correlation between fractal time series, the more similar are their fractal characteristics.

## VI. CONCLUSION

In the work, time series correlation analysis was carried out revealed a significant correlation between the Bitcoin rate and the activity of the social communities associated with it. Time series fractal analysis indicated the presence of self-similar and multifractal properties, both for the series of Bitcoin prices, and for ones of community activities.

The results of the study have shown that the series with a correlation dependence have a similar multifractal structure. Namely, the greater correlation between fractal series, the more similar are their fractal characteristics.

Thus, the fractal structure of time series of cryptocurrency and activities of social communities, and, consequently, the dynamics of their development, are interrelated. This allows predicting the Bitcoin crypto currency rate based on data derived from social networks.

Our future research will focus on study based on the multifractal analysis of the degree of social networks influence on the financial dynamics. The research results can be used by professionals in the field of analysis and investment.

REFERENCES

[1] Peng Xie, Hailiang Chen and Yu Jeffrey Hu. "Network Structure and Predictive Power of Social Media for the Bitcoin Market", *Georgia Tech Scheller College of Business Research* Paper No. 17-5, 2017. Available: https://ssrn.com/abstract=2894089.

[2] Do, Rotimi Opeke and James Webb. *CS224W Final Project: Predicting Yelp Ratings From Social Network Data*, Department of Computer Science, Stanford University, p.8, December 9, 2015.

[3] Nakamoto Satoshi. *Bitcoin: A Peer-to-Peer Electronic Cash System*, 31 October 2008.

[4] Edgar E. Peters.*Fractal Market Analysis: Applying Chaos Theory to Investment and Economics*, John Wiley & Sons, 1994.

[5] Guillermo Romero Meléndez. *The fractal nature of bitcoin: evidence from wavelet power spectra*, Fundacion Universidad de las Americas Puebla, 2014.

[6] Sho Tsugawa and Hiroyuki Ohsaki. "Emergence of Fractals in Social Networks: Analysis of Community Structure and Interaction Locality", in *38th Annual Computer Software and Applications Conference*, 2014.

[7] Valerio Arnaboldi, Andrea Passarella, Marco Conti and Robin I.M. Dunbar. *Online Social Networks: Human Cognitive Constraints in Facebook and Twitter Personal Graphs*, Elsevier, 2015.

[8] Alex Greaves and Benjamin Au. "Using the Bitcoin Transaction Graph to Predict the Price of Bitcoin", *Department of Computer Science*, Stanford University, p.8, December 8, 2015.

[9] Martina Matta. *The Predictor Impact of Web Search and Social Media*, Theses doctoral research. University of Cagliari, 2016.

[10] A. Mittal and A. Goel. "Stock Prediction Using Twitter Sentiment Analysis" in *Proceeding of IEEE/WIC/ACM International Conference on Web Intelligence and Intelligent Agent Technology*, 2013, p.1-5.

[11] Bollen Johan, Huina Mao, and Xiaojun Zeng. "Twitter mood predicts the stock market." *Journal of Computational Science*, vol. 2.1, p.1-8, 2011.

[12] Feng Mai, Qing Bai, Zhe Shan, Xin (Shane) Wang and Roger H. L. Chiang. "From Bitcoin to Big Coin: The Impact of Social Media on Bitcoin Performance", 2016. Available: http://ssrn.com/abstract=2545957

[13] D Garcia, C.J Tessone, P Mavrodiev and N Perony. "The digital traces of bubbles: feedback cycles between socio-economic signals in the Bitcoin economy", *J. R. Soc. Interface* 11, 2014. Available: http://dx.doi.org/10.1098/rsif.2014.0623

[14] J.W. Kantelhardt. "Fractal and multifractal time series", *Mathematics of complexity and dynamical systems*, pp. 463-487, 2012.

[15] R.H.Riedi. "Multifractal processes", in Doukhan P., Oppenheim G., Taqqu M.S. (Eds.), *Long Range Dependence: Theory and Applications:* Birkhuser, 2002, pp. 625–715.

[16] J.W. Kantelhardt, E. Koscielny-Bunde, H.H.A. Rego, S. Havlin and A. Bunde. "Detecting long-range correlations with detrended fluctuation analysis", *Physica A: Statistical Mechanics and its Applications*, vol. 295, pp. 441-454, 2001.

[17] D.V. Ageyev and A.N. Kopylev, "Modelling of multiservice streams at the decision of tasks of parametric synthesis," in *2013 23rd International Crimean Conference "Microwave & Telecommunication Technology"*, Sevastopol, 2013, pp. 505-506.

[18] D. Ageyev and N. Qasim, "LTE EPS network with self-similar traffic modeling for performance analysis," *2015 Second International Scientific-Practical Conference Problems of Infocommunications Science and Technology (PIC S&T)*, Kharkiv, 2015, pp. 275-277.

[19] D.V. Ageyev and M.T. Salah. "Parametric synthesis of overlay networks with self-similar traffic", *Telecommunications and Radio Engineering*, no.75(14), pp. 1231-12141, 2016.

[20] D. Ageyev and D. Evlash. "Multiservice telecommunication systems design with network's incoming self-similarity flow", in *2008 Proceedings of International Conference on Modern Problems of Radio Engineering, Telecommunications and Computer Science*, 2008, pp. 403-405.